\documentclass{article}

\usepackage{amsmath,amsfonts,amssymb,xypic,xy,graphicx,caption,hyperref, graphics,epsfig, epstopdf}
\usepackage{fullpage,setspace,color}

\begin{document}
\title{\normalsize \textbf{Investigation of the Gravitational Potential Dependence of the Fine-Structure Constant Using Atomic Dysprosium}}
\author{\normalsize S.J. Ferrell,$^1$ A. Cing\"{o}z,$^1$ A. Lapierre,$^2$ A.-T. Nguyen,$^3$ N. Leefer,$^1$ D. Budker,$^{1,4}$ \\ \normalsize V.V. Flambaum$^{5,6}$, S.K. Lamoreaux,$^{7}$, and J.R. Torgerson$^3$ \\ \footnotesize \emph{$^1$Department of Physics, University of California at Berkeley, Berkeley, California 94720-7300, USA} \\ \footnotesize \emph{$^2$TRIUMF National Laboratory, 4004 Wesbrook Mall, Vancouver, British Columbia, V6T 2A3, Canada} \\ \footnotesize \emph{$^3$Physics Division, Los Alamos National Laboratory, P-23, MS-H803, Los Alamos, New Mexico 87545, USA} \\ \footnotesize \emph{$^4$Nuclear Science Division, Lawrence Berkeley National Laboratory, Berkeley, California 94720, USA} \\ \footnotesize \emph{$^5$ School of Physics, The University of New South Wales, Sydney NSW 2052, Australia} \\ \footnotesize \emph{$^6$ Institute for Advanced Study, Massey University (Albany Campus), Private Bag 102904, North Shore MSC Auckland, New Zealand}\\ \footnotesize \emph{$^7$ Department of Physics, Yale University, New Haven, Connecticut 06520-8120, USA}}
\date{}
\maketitle
\begin{abstract}
Radio-frequency E1 transitions between nearly degenerate, opposite
parity levels of atomic dysprosium were monitored over an eight
month period to search for a variation in the fine-structure
constant.  During this time period, data were taken at different
points in the gravitational potential of the Sun.  The data are
fitted to the variation in the gravitational potential yielding a
value of $(-8.7 \pm 6.6) \times 10^{-6}$ for the fit parameter
$k_\alpha$. This value gives the current best laboratory limit. In
addition, our value of $k_{\alpha}$ combined with other experimental
constraints is used to extract the first limits on $k_e$ and $k_q$.
These coefficients characterize the variation of $m_e/m_p$ and
$m_q/m_p$ in a changing gravitational potential, where $m_e$, $m_p$,
and $m_q$ are electron, proton, and quark masses. The results are
$k_e = (4.9 \pm 3.9) \times 10^{-5}$ and $k_q = (6.6 \pm 5.2) \times
10^{-5}$.
\end{abstract}

\begin{figure*}[h!]
\begin{center}
\includegraphics[width=3in, height=2.5in]{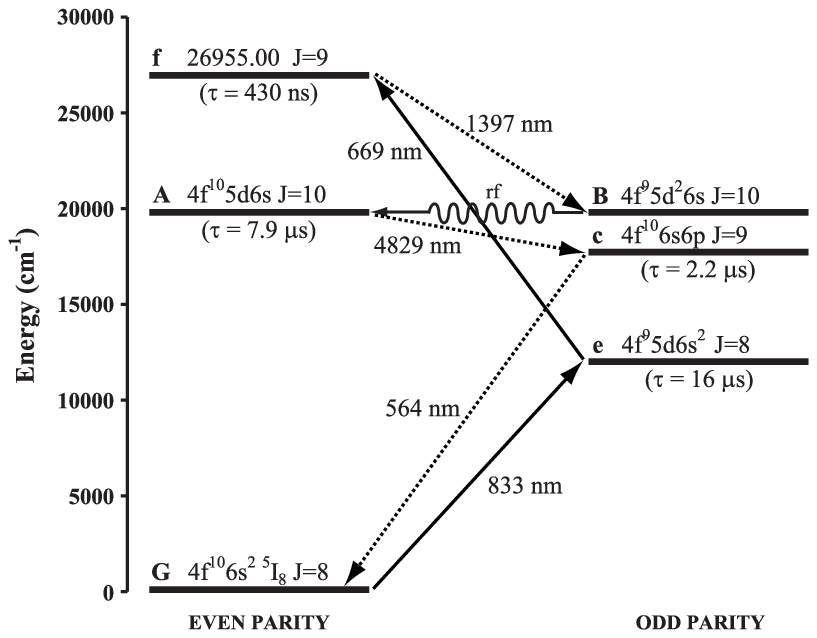}
\caption{The relevant energy-level, population, and detection scheme for the dysprosium experiment. \label{levels}}
\end{center}
\end{figure*}

\noindent According to general relativity, values of fundamental
constants are independent of space and time as required by the
Einstein Equivalence Principle (EEP).  Modern theories that aim to
unify gravitation with the other forces, however, do not set this
restriction on the fundamental constants; so, they may vary either
spatially or temporally, which is in conflict with EEP \cite{1}.
\\
\\
Various recent studies have reported the results of searches for a
temporal variation of the fine-structure constant, $\alpha = e^2 /
\hbar c$.  These include the analyses of absorption spectra from
quasars [2-5], the analyses of the nuclear products of the natural
fission reactor at Oklo, which operated $2\times10^9$ years ago
[6-9], and various laboratory investigations involving atomic
clocks, see for example Refs. [10-13].  In contrast to studies
involving analyses of the processes that have occurred billions of
years ago, laboratory searches are sensitive to present-day
variation of fundamental constants.  Recently, our group's
experiment utilizing the $E1$ radio frequency (rf) transitions
between nearly degenerate opposite parity levels in atomic
dysprosium has yielded a result of $\dot{\alpha}/\alpha = (-2.7 \pm
2.6) \times 10^{-15}$ /yr \cite{14} over an eight month observation
period.  A detailed description of the experimental setup and
analysis was given in Refs. [14-16].  Here,  we investigate a
possible correlation between a change in $\alpha$ and a change in
the gravitational potential.
\\
\\
\noindent The energy of an atomic level can be written as
\begin{equation}
E = h\nu = E_0 + q\left(\frac{\alpha^2}{\alpha_0^2} -1\right) \mbox{ ,}
\end{equation}
where $E_0$ and $\alpha_0$ are the present-day values of the energy and $\alpha$, respectively, and $q$ contains the sensitivity of the level to $\alpha$ \cite{17}.   From Eq. (1), a change in $\alpha$ should result in a change in  $\nu$, for $\alpha \approx \alpha_0$, as given by
\begin{equation}
h\delta\nu = 2q\frac{\delta\alpha}{\alpha}\mbox{ .}
\end{equation}
Similarly, the change of the rf transition frequency between levels $A$ and $B$, as shown in Fig. \ref{levels}, is given by
\begin{equation}
\delta(\Delta\nu) = 2\frac{\left(q_B - q_A\right)}{h}\frac{\delta\alpha}{\alpha}\mbox{ ,}
\end{equation}
where $|2(q_B - q_A) / h| \approx 1.8 \times 10^{15}$~Hz \cite{17}.
Here, we take this number to be exact, but the uncertainty is
expected to be approximately twenty percent.
\\
\\

\begin{figure*}[h!]
\begin{center}
\includegraphics[width=3in, height=2.5in]{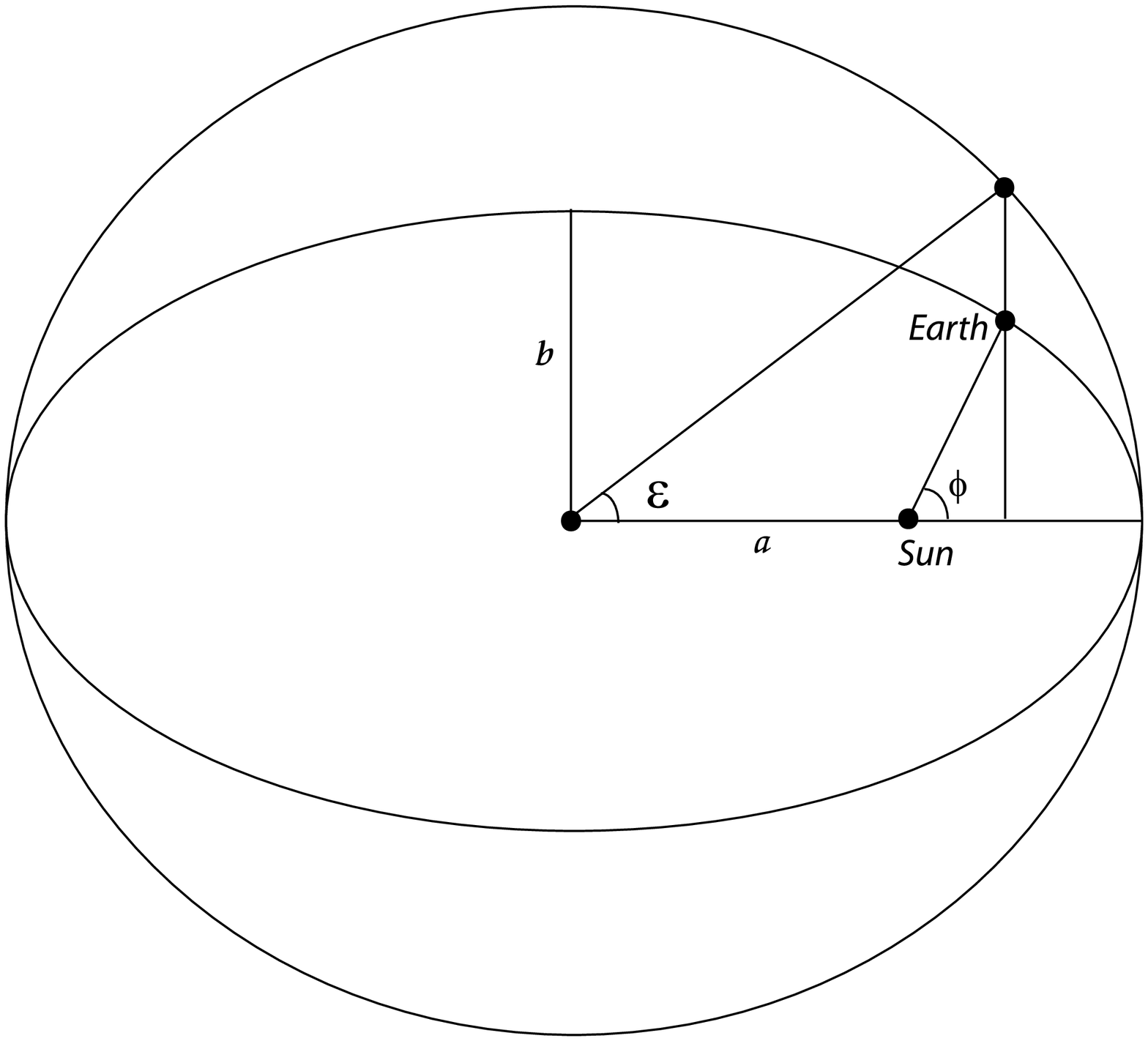}
\includegraphics[width=2.3in, height=3in, angle=90]{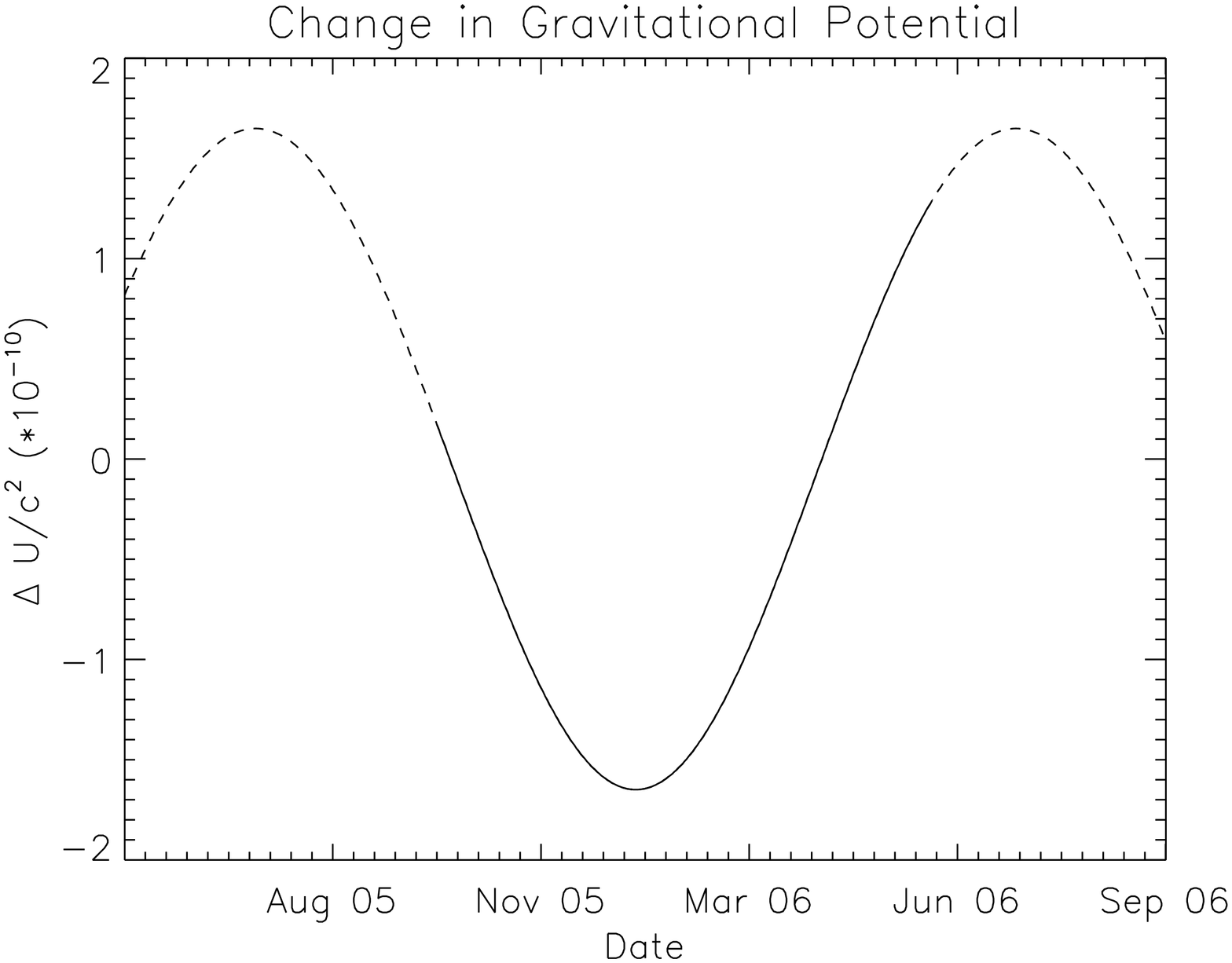}
\caption{Left:  The true anomaly, $\phi$, is the angle subtended from perihelion,
 the point of closest approach.  The eccentric anomaly, $\varepsilon$, is the angle
 between perihelion and the position of the Earth in its orbit projected onto the
 auxiliary circle of the ellipse (the eccentricity of the ellipse is exaggerated for clarity). Right:  The change in gravitational potential of the Sun at the Earth due to the ellipticity of the orbit. The solid line represents the time period during which data were taken.  \label{grav}}
\end{center}
\end{figure*}

\noindent The transitions utilized in the experiment are the 235-MHz
(J = 10 $\rightarrow$ J = 10) transition of the isotope $^{162}$Dy
and the 3.1-MHz (F = 10.5 $\rightarrow$ F = 10.5) transition of
$^{163}$Dy.  The long-lived level $B$ (lifetime $\tau>200\ \mu$s
\cite{18}) is populated via three transitions.  Laser light at 833
nm and 669 nm is used for the first two transitions, and then the
atoms spontaneously decay to level $B$ as shown in Fig.
\ref{levels}. The transition between levels $A$ and $B$ is induced by
an applied rf electric field.  The atoms decay from level $A$ in two steps,
 and the fluorescence at 564 nm is monitored to detect the rf transition. \\
\\
\noindent Data were taken over an eight month period and hence at
different points in the gravitational potential of the Sun.  The
gravitational potential at the Earth due to the Sun \cite{19} is
given by
\begin{equation}
U = \frac{-GM_s}{r}\mbox{ ,}
\end{equation}
where $G$ is the gravitational constant and $M_s$ is the mass of the
Sun.  For an elliptical orbit, the distance $r$ between the Sun and
Earth is
\begin{equation}
r = a\frac{1 - \epsilon^2}{1 + \epsilon\cos\phi}\mbox{ ,}
\end{equation}
where $a$ is the semi-major axis of the Earth's orbit, $\epsilon =
\sqrt{1 - b^2/a^2} \approx 0.0167$ is the eccentricity, $b$ is the
semi-minor axis, and $\phi$ is the true anomaly (see Fig.
\ref{grav}).  Substituting Eq. (5) in to Eq. (4), the gravitational
potential becomes
\begin{equation}
U = \frac{-GM_s}{a} - \frac{GM_s}{a}\,\epsilon\cos\phi +
O(\epsilon^2)\mbox{ .}
\end{equation}

\begin{figure*}[h!]
\begin{center}
\includegraphics[width=4in, height=6in, angle=90]{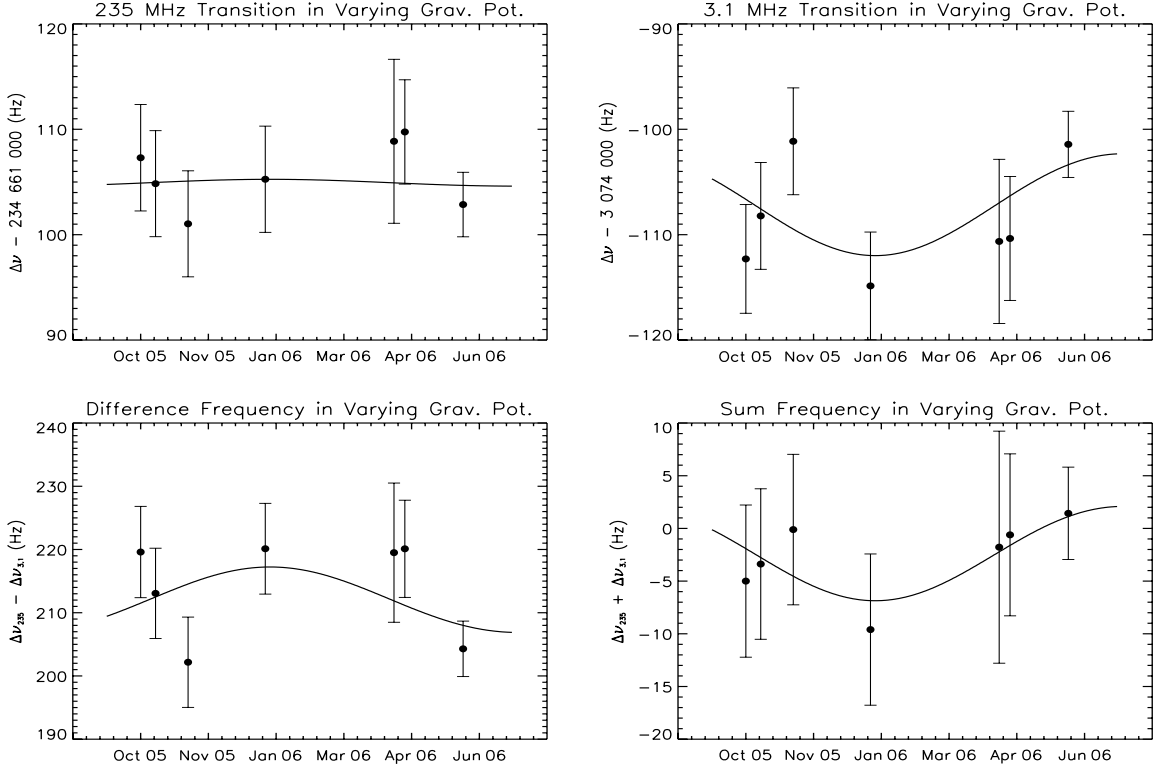}
\caption{Top left: The data fitted to the gravitational potential for the 235-MHz transition.  Top right: The data fitted to the gravitational potential for the 3.1-MHz transition.   The bottom plots show the difference and sum frequencies fitted to the change in gravitational potential.  \label{fits}}
\end{center}
\end{figure*}

\noindent The first term in Eq. (6) is the constant part of the
gravitational potential at Earth, while the second term is a change
in gravitational potential, $\Delta U$, which arises due to the
eccentricity of the Earth's orbit.  The fractional change in
$\alpha$ as a function of gravitational potential can be
parametrized as \cite{20}
\begin{eqnarray}
\frac{\delta\alpha}{\alpha} &=& k_{\alpha} \frac{\Delta U(t)}{c^2}\mbox{ ,} \\
\Delta U(t) &=& - \frac{GM_s}{a}\,\epsilon\cos\phi(t)\mbox{ .}
\end{eqnarray}
The data were taken between October 2005 to June 2006; the true
anomaly is zero at the 2005 perihelion, January 2 \cite{21}.  The
true anomaly was calculated for each data point using a two step
process.  The elapsed time in days from perihelion is related to the
eccentric anomaly (see Fig. \ref{grav}) as \cite{22}
\begin{equation}
t = \sqrt{\frac{a^3}{GM}}\left(\varepsilon - \epsilon\sin\varepsilon\right)\mbox{ ,}
\end{equation}
where $M = M_s + M_E$ is the sum of the masses of the Sun and Earth.
The eccentric anomaly is obtained by solving Eq. (9) and then used
to calculate the cosine of the true anomaly as a function of time
since
\begin{equation}
\cos\phi = \frac{\cos\varepsilon - \epsilon}{1 - \epsilon\cos\varepsilon}\mbox{ .}
\end{equation}
The calculated values for $\cos\phi$ are substituted into Eq. (8) to
find the gravitational potential for each data point.  The measured
frequencies for each isotope are fitted to the gravitational
potential using a two-parameter least squares fit given by
\begin{equation}
\delta(\Delta\nu) - \nu^*= x_0\frac{\Delta U(t)}{c^2} + x_1\mbox{ ,}
\end{equation}
where $\nu^*$ is an arbitrary reference frequency, and $x_0$ and
$x_1$ are the fit parameters. The parameter $x_1$ accounts for the
offset due to the reference frequency while the parameter $x_0$
determines the correlation between the change in gravitational
potential and the transition frequency. If such a correlation is due
to a change in $\alpha$, the sum and difference frequencies may also
be fit to the varying potential to extract additional information
regarding the variation of $\alpha$. Since the energy difference
between levels $A$ and $B$ is of opposite sign for the two isotopes
and the $q$ values are independent of the nucleus, the sum of the
transition frequencies should be insensitive to $\alpha$ variation,
while the difference of the transition frequencies should be twice
as sensitive [14,15]. The data plots and the fits of the measured
frequencies for each isotope as well as the sum and difference
frequencies are shown in Fig. \ref{fits}.

\noindent The parameter $x_0$ obtained by the least-squares fit is
$(-0.2 \pm 1.7) \times 10^{10}$~Hz for the 235-MHz transition, $(2.9
\pm 1.7) \times 10^{10}$~Hz for the 3.1-MHz transition, $(2.7 \pm
2.4) \times 10^{10}$~Hz for the sum frequency, and  $(-3.1 \pm 2.4)
\times 10^{10}$~Hz for the difference frequency.
\\
\\
\noindent These values can be used to calculate the constraint on
the parameter $k_{\alpha}$ from Eq. (7).  Substituting Eq. (3) into
Eq. (7), we get a relation similar to Eq. (11),
\begin{equation}
\delta(\Delta\nu) = \left(2\frac{q_B - q_A}{h}\right)k_{\alpha} \frac{\Delta U(t)}{c^2}\mbox{ .}
\end{equation}

\noindent The fit parameter yields $k_{\alpha} = (-1.1 \pm 9.2)
\times 10^{-6}$ for the 235-MHz transition, $k_{\alpha} = (-16.3 \pm
9.4) \times 10^{-6}$ for the 3.1-MHz transition, $k_{\alpha} = (-8.7
\pm 6.6) \times 10^{-6}$ for the difference frequency, and $(-7.5
\pm 6.6) \times 10^{-6}$ for the sum frequency. Although the value
for the sum frequency is expected to be consistent with zero, the
one-sigma mismatch of our value is consistent with the estimated
uncertainties, dominated by systematic effects \cite{14}.
\\
\\
\noindent The results relating a changing $\alpha$ to a changing
gravitational potential are summarized in Table \ref{results}.  The
results in the first row of Table \ref{results} depend on the
variation of $\alpha$ as well as the variation of $m_e/m_p$. The
dependence on $m_e / m_p$ is contained in $k_e$ \cite{20}. The
results in the second and third rows in Table \ref{results} depend
on the variation of $\alpha$ and the variation of $m_q/m_p$, which
is contained in $k_q$ \cite{20}.  In addition to the results shown
in Table \ref{results}, our results may limit the parameter range in
some specific theories, such as those in Refs. [23-25].
\\
\\
Our calculated value of $k_{\alpha}$ can be combined with the
results for $k_{\alpha} + 0.17k_e$ and $k_{\alpha} + 0.13k_q$ to
extract a value of $(4.9 \pm 3.9) \times 10^{-5}$ for $k_e$ and
$(6.6\pm 5.2) \times 10^{-5}$ for $k_q$.

\begin{table*}[h!]
\begin{center}
\begin{tabular}{c|c|c}
Parameter & Constraint & Experimental Ref. \\ \hline \hline
$k_{\alpha} + 0.17k_e$ & $(-3.5 \pm 6) \times 10^{-7}$ & \cite{26}
\\ \hline $|k_{\alpha} + 0.13k_q|$ & $< 2.5 \times 10^{-5}$ &
\cite{27}  \\ \hline $k_{\alpha} + 0.13k_q$ & $(-1 \pm 17) \times
10^{-7}$ & \cite{28} \\ \hline $k_{\alpha}$ & $(-8.7 \pm 6.6) \times
10^{-6}$ & this work \\ \hline $k_e$ & $(4.9 \pm 3.9) \times
10^{-5}$ & this work \\ \hline $k_q$ & $(6.6 \pm 5.2) \times
10^{-5}$ & this work \\ \hline
\end{tabular}
\caption{A summary of results for changing fundamental constants in
a varying gravitational potential based on the theoretical work from
Ref. \cite{20}. \label{results}}
\end{center}
\end{table*}

\noindent In this paper, we have reported the first laboratory
result for $k_{\alpha}$ and subsequently extracted a limit on $k_e$
and $k_q$.  In principle the results from Ref. \cite{26} can also be
combined with other optical-Cs clock comparisons in such a way as to
extract $k_{\alpha}$ independent of $k_e$.  Currently, our results
are limited by systematic uncertainties \cite{14}.  A new apparatus
is under construction to address the systematic effects and, thus,
possibly increase the sensitivity of the experiment by up to three
orders of magnitude \cite{15}.
\\
\\
The authors are grateful to M. Kozlov for stimulating discussions.
This work has been supported in part by  Los Alamos National
Laboratory LDRD, NSF REU supplement, and by grant RFP1-06-15 from
the Foundational Questions Institute (fqxi.org).


\begin{thebibliography}{1}
\bibitem{1} J.-P. Uzan, Rev. Mod. Phys. \textbf{75}, 403 (2003).
\bibitem{2} J.K. Webb \emph{et al}., Phys. Rev. Lett. \textbf{87}, 091301 (2001).
\bibitem{3} M.T. Murphy \emph{et al}., Mon. not. R. Astron. Soc. \textbf{345}, 609 (2003).
\bibitem{4} R. Quast \emph{et al}., Astron. Astrophys. \textbf{415}, L7 (2004).
\bibitem{5} R. Srianand, H. Chand, P. Petitjean, and B. Aracil, Phys. Rev. Lett. \textbf{92} 121302 (2004).
\bibitem{6} T.Damour and F. Dyson, Nucl. Phys. \textbf{B480}, 37 (1996).
\bibitem{7} Y. Fujii \emph{et al}., Nucl. Phys. \textbf{B573}, 377 (2000).
\bibitem{8} S.K. Lamoreaux and J.R. Torgerson, Phys. Rev. D \textbf{69}, 121701(R) (2004).
\bibitem{9} C.R. Gould, E.I. Sharapov, and S.K. Lamoreaux, Phys. Rev. C \textbf{74}, 024607 (2006).
\bibitem{10} S. Bize \emph{et al}., Phys. Rev. Lett. \textbf{90}, 150802 (2003).
\bibitem{11} H. Marion \emph{et al}., Phys. Rev. Lett. \textbf{90}, 150801 (2003).
\bibitem{12} M. Fischer \emph{et al}., Phys. Rev. Lett. \textbf{92} 230802 (2004).
\bibitem{13} E. Peik \emph{et al}., Phys. Rev. Lett. \textbf{93}, 170801 (2004).
\bibitem{14} A. Cing\"{o}z \emph{et al}., Phys. Rev. Lett. \textbf{98}, 040801 (2007).
\bibitem{15} A.T. Nguyen, D. Budker, S.K. Lamoreaux, and J.R. Torgerson, Phys. Rev. A \textbf{69}, 022105 (2004)
\bibitem{16} A. Cing\"{o}z \emph{et al}., Phys. Rev. A \textbf{72}, 063409 (2005).
\bibitem{17} V.A. Dzuba, V.V. Flambaum, and M.V. Marchenko, Phys. Rev. A \textbf{68}, 022506 (2003).
\bibitem{18} D. Budker, D. DeMille, E.D. Commins, and M.S. Zolotorev, Phys. Rev. A \textbf{50}, 132-143 (1994).
\bibitem{19} The variation of the gravitational potential due to Jupiter and the other planets was ignored since it is at least two orders of magnitude smaller than that due to the Sun.
\bibitem{20} V.V. Flambaum, arXiv:0705.3704 physics.atom-ph, \emph{to be published in} Int. J. Mod.
Phys. A
\bibitem{21} http://aa.usno.navy.mil/data/docs/EarthSeasons.html
\bibitem{22} L. Hand and J. Finch, \underline{Analytical Mechanics}, Cambridge University Press 1998.
\bibitem{23} J. Magueijo, J.D. Barrow, and H.B. Sandvik, Phys. Lett. B \textbf{549}, 284-289 (2002).
\bibitem{24} H.B. Sandvik, J.D. Barrow, and J. Magueijo, Phys. Rev. Lett. \textbf{88}, 031302 (2002).
\bibitem{25} J. Magueijo,Phys. Rev. D \textbf{62}, 103521 (2000).
\bibitem{26} Fortier \emph{et al}., Phys. Rev. Lett. \textbf{90}, 070801 (2007).
\bibitem{27} A. Bauch and S. Weyers, Phys. Rev. D \textbf{65}, 081101R (2002).
\bibitem{28} N. Ashby \emph{et al}. Phys. Rev. Lett. \textbf{98}, 070802 (2007).
\end{thebibliography}
\end{document}